\documentclass[11pt,english,epsf]{scrartcl}
\usepackage{lmodern}
\usepackage[T1]{fontenc}
\usepackage[latin9]{inputenc}
\usepackage{geometry}
\usepackage{amsmath}
\usepackage{amssymb}

\newtheorem{lemma}{Lemma}
\newcommand {\dfn} {\stackrel{\Delta} {=}}
\newcommand {\exe} {\stackrel{\cdot} {=}}

\newcommand {\reals} {{\rm I\!R}}

\newcommand {\bx} {\mbox{\boldmath $x$}}

\newcommand {\bE} {\mbox{\boldmath $E$}}

\newcommand {\bX} {\mbox{\boldmath $X$}}

\newcommand{\calD}{{\cal D}}
\newcommand{\calE}{{\cal E}}

\newcommand{\calH}{{\cal H}}
\newcommand{\calI}{{\cal I}}

\newcommand{\calT}{{\cal T}}

\newcommand{\calX}{{\cal X}}

\allowdisplaybreaks

\begin{document}
\thispagestyle{empty}
\title{Asymptotically Optimal Decision Rules for Joint Detection and Source
Coding
}
\author{Neri Merhav
}
\date{}
\maketitle

\begin{center}
Department of Electrical Engineering \\
Technion - Israel Institute of Technology \\
Technion City, Haifa 32000, ISRAEL \\
E--mail: {\tt merhav@ee.technion.ac.il}\\
\end{center}
\vspace{1.5\baselineskip}
\setlength{\baselineskip}{1.5\baselineskip}

\begin{center}
{\bf Abstract}
\end{center}
\setlength{\baselineskip}{0.5\baselineskip}
The problem of joint detection and lossless source coding
is considered. We derive asymptotically optimal decision rules
for deciding whether or not a sequence of observations has emerged from a desired
information source, and to compress it if has. In particular, our decision rules
asymptotically minimize the cost of compression in the
case that the data has been classified as `desirable', subject to given
constraints on the two kinds of the probability of error. In another version of
this performance criterion, the constraint on the false alarm probability is
replaced by the a constraint on the cost of compression in the false alarm
event. We then analyze the asymptotic performance of these decision rules and
demonstrate that they may exhibit certain phase transitions. We also derive
universal decision rules for the case where the underlying
sources (under either hypothesis or both) are unknown, and
training sequences from each source may or may 
not be available. Finally, we discuss how our
framework can be extended in several directions.\\

\vspace{0.2cm}


\setlength{\baselineskip}{2\baselineskip}
\newpage

\section{Introduction}

Classical hypothesis testing theory, based on the Neyman--Pearson theorem
(see, e.g., \cite[Sect.\ 11.7]{CT06}),
provides the optimal rule for deciding between two hypotheses concerning the
distribution or density of a given observation or sequence of observations. It tells us
that best trade-off between the two kinds of probability of error is achieved
by the likelihood ratio test.

In certain situations, however, this decision between the two hypotheses might be
only one of the tasks to be carried out. For example, consider a scenario
where under hypothesis $\calH_0$, the sequence of observations that we receive
is just pure noise, which contains no useful information that may interest us, whereas under
hypothesis $\calH_1$, the data that we have at hand has emerged from a desirable
information source, and in this case, further processing is called for, such as
lossless or lossy data compression, parameter estimation \cite{Moustakides11},
\cite{MJTW12}, channel decoding \cite{Merhav13}, \cite{Wang10}, \cite{WCCW11},
encryption, further classification, etc.

The straightforward
approach to this problem would be to first apply Neyman--Pearson hypothesis
testing, and then,
if hypothesis $\calH_1$ is accepted, perform the corresponding task
using the best strategy available. This approach {\it separates} between
optimal decision and the optimality of the subsequent task.
A more sophisticated approach, however, is to solve the two problems jointly, namely,
to devise a decision rule that takes into account also the cost of the
subsequent task (in case it is to be carried out), and on the other
hand, optimize the strategy of the following task, taking into account that the data belongs to
the decision region of $\calH_1$. 

For the case where the second task is
Bayesian parameter estimation, Moustakides \cite{Moustakides11}
and Moustakides {\it et al.} \cite{MJTW12} have derived an optimal solution
for the combined problem.
In particular, in these articles, the problem of joint detection and estimation
was posed and solved under the criterion of 
minimizing the conditional expected cost of the estimation error,
given that the data is classified into $\calH_1$ subject to certain
constraints on the false alarm (FA) and misdetection (MD) probabilities
(or related constraints). The optimal decision rule, under this criterion,
is interesting, but it turns out to be rather complicated and non--trivial in three respects: (i)
the proof of optimality is quite long and not easy, (ii) the insight behind
this decision rule is not obvious, and (iii) it may be difficult to implement.

In this paper, we propose a modified criterion,\footnote{Details will
follow in the sequel.} which is
asymptotically equivalent for a large number of observations, at least in the
relevant regime, where the MD probability is constrained to tend to zero.
The point of this modification in the criterion is that it 
allows us to use a slightly extended
version of the
Neyman--Pearson lemma in order to
derive the optimal decision rule in a fairly simple and easy manner. It is also
rather easy to implement, or at least to approximate by an easily implementable
decision rule. Finally, the intuition behind this decision rule is easier to
grasp. We focus, in this paper, on memoryless sources and on the case where the second task to be
performed, after the detection, is lossless data compression, but this should
be considered only as an example,
as the methodology proposed is applicable for a wide
variety of tasks, as will be discussed. In fact, the same methodology has already been used in
\cite{Merhav13}, where under $\calH_1$, the observed data is the output of a
noisy channel fed by a codeword, and the second task after detection is channel decoding,
with application to (slotted) asynchronous communication (see also 
\cite{Wang10} and \cite{WCCW11} for earlier work).

In addition to the derivation of the optimal decision rule under our modified
criterion, we also analyze its performance in terms of asymptotic
exponents. One of our findings is that these asymptotic exponents may exhibit
``phase transitions'' in the sense of having discontinuous derivatives as
functions of the parameters of the problem. Such phase transitions do not occur
in the ordinary Neyman--Pearson decision rule.
Finally, we derive universal
versions of our decision rule that are suitable for scenarios where at least
one of the probability distributions (under $\calH_0$ and/or $\calH_1$) is
unknown (yet they are still known to be memoryless), 
and we might have access to a training sequence from one of the
sources or both.
We also discuss, as mentioned earlier, how our method applies to tasks other than lossless source
coding as well as more general classes of sources.

The outline of the remaining part of this paper is as follows.
In Section 2, we establish notation conventions and define the problem
in several different versions.
In Section 3, we present the above--mentioned extension of the Neyman--Pearson
lemma. In Section 4, we apply this lemma to
the solution of one version of the joint detection and compression problem,
and in Section 5 we analyze its performance and discuss it.
In Section 6, we show how to apply Lemma 1 to a number of other variants of the problem.
Section 7 is devoted to universal decision rules, and finally,
in Section 8 we conclude.

\section{Notation Conventions and Problem Formulation}

Throughout the paper, random variables will be denoted by capital
letters, specific values they may take will be denoted by the
corresponding lower case letters, and their alphabets
will be denoted by calligraphic letters. Random
vectors and their realizations will be denoted,
respectively, by capital letters and the corresponding lower case letters,
both in the bold face font. Their alphabets will be superscripted by their
dimensions. For example, the random vector $\bX=(X_1,\ldots,X_n)$, ($n$ --
positive integer) may take a specific vector value $\bx=(x_1,\ldots,x_n)$
in $\calX^n$, the $n$--th order Cartesian power of $\calX$, which is
the alphabet of each component of this vector. In this paper, $\bX$ emerges
from either one of two sources, $P_0$ or $P_1$. The probability of an event
$\calE$ under $P_i$ will be denoted by $P_i(\calE)$  and the expectation
operator w.r.t.\ $P_i$ will be denoted by $\bE_i\{\cdot\}$, $i=0,1$.
The entropy of a generic distribution $Q$ on $\calX$ will be denoted by
$H(Q)$. The notation $H(P_i)$ will be shortened to $H_i$, $i=0,1$.
For two positive sequences $a_n$ and $b_n$, the notation $a_n\exe b_n$ will
stand for equality in the exponential scale, that is,
$\lim_{n\to\infty}\frac{1}{n}\log \frac{a_n}{b_n}=0$. The indicator function
of an event $\calE$ will be denoted by $\calI\{E\}$.
The empirical distribution of a sequence $\bx\in\calX^n$, which will be denoted
by $\hat{P}_{\bx}$, is the vector of relative frequencies $\hat{P}_{\bx}(x)$ of
each symbol $x\in\calX$ in $\bx$.
The type class of $\bx\in\calX^n$, denoted $\calT_{\bx}$, is the set of all vectors $\bx'$ 
with $\hat{P}_{\bx'}=\hat{P}_{\bx}$. When we wish to emphasize the
dependence of the type class on the empirical distribution $\hat{P}$, we denote it by
$\calT(\hat{P})$.

Let $\bX=(X_1,\ldots,X_n)$ be a sequence of random variables 
drawn from a finite alphabet memoryless source. There are two hypotheses
concerning the probability distribution of the underlying source: Under
hypothesis $\calH_0$, the source is $P_0=\{P_0(x),~x\in\calX\}$, whereas under
hypothesis $\calH_1$, the source is $P_1=\{P_1(x),~x\in\calX\}$.
The source $P_0$ designates unwanted data
(e.g., pure noise, spam, meaningless or unimportant data), 
while the source $P_1$ represents useful, desirable information,
which we would like to keep for further processing. In this paper, this
further processing
is lossless data compression (source coding). 

A decision rule is a
partition of $\calX^n$, the space of source vectors of length $n$, into two
complementary regions $\Omega\subseteq\calX^n$ and
$\Omega^c=\calX^n\setminus\Omega$, where $\Omega$ is the region where we
accept $\bX$ as having emerged from $P_1$,
and $\Omega^c$ is the region where we classify it as having been generated by $P_0$.
Thus, only source vectors that fall in $\Omega$ are to be compressed.
Since the decision rule is fully defined by the choice of the subset $\Omega$, we will
sometimes use expressions like ``the decision rule $\Omega$'' as shorthand for  ``the
decision rule associated with $\Omega$,'' with a slight abuse of formal
preciseness. 

Our aim is to find a decision rule and a compression strategy 
that jointly optimize the compression performance
within $\Omega$ subject to constraints on the error probabilities of the two
kinds: $P_0(\Omega)$ -- the probability of false alarm (FA), and
$P_1(\Omega^c)$ -- the probability of misdetection (MD).
In particular, let $L:\calX^n\to \{0,1,2,\ldots\}$ be a
length function of a lossless fixed--to--variable length code that satisfies Kraft's inequality
\begin{equation}
\sum_{\bx\in\calX^n}2^{-L(\bx)}\le 1.
\end{equation}
A seemingly natural goal (in the spirit of \cite{MJTW12})
would be to solve the problem:
\begin{eqnarray}
\label{min}
& &\mbox{minimize}~~~\bE_1\{L(\bX)|\bX\in\Omega\}\\
& &\mbox{subject to}~~P_0(\Omega)\le \epsilon_{\mbox{\tiny
FA}}\nonumber\\
& &~~~~~~~~~~~~~~~P_1(\Omega^c)\le \epsilon_{\mbox{\tiny MD}}\nonumber
\end{eqnarray}
where the minimization is over the length function $L(\cdot)$ and the choice
of $\Omega$, and
where $\epsilon_{\mbox{\tiny FA}}$ and
$\epsilon_{\mbox{\tiny MD}}$ are prescribed numbers designating the maximum
tolerable FA and MD probabilities, respectively. Of course,
$\epsilon_{\mbox{\tiny FA}}$ and $\epsilon_{\mbox{\tiny MD}}$ should not be
chosen both too small, otherwise, the two constraints may become contradictory
(the minimum achievable $\epsilon_{\mbox{\tiny MD}}$ for a given
$\epsilon_{\mbox{\tiny FA}}$ is achieved by the performance of the ordinary likelihood ratio
test).

Now, it makes sense to let
$\epsilon_{\mbox{\tiny MD}}$ and $\epsilon_{\mbox{\tiny FA}}$ decay
exponentially with $n$. We let then $\epsilon_{\mbox{\tiny
MD}}=\exp(-nE_{\mbox{\tiny MD}})$ and 
$\epsilon_{\mbox{\tiny
FA}}=\exp(-nE_{\mbox{\tiny FA}})$, where $E_{\mbox{\tiny MD}}$ and
$E_{\mbox{\tiny FA}}$ are positive constants, independent of $n$.
In this regime, $P_1(\Omega)\ge 1-\exp(-nE_{\mbox{\tiny MD}})$ tends to unity,
and so, the conditioning on $\bX\in\Omega$, that appears in the objective
function of (\ref{min}) has an asymptotically vanishing effect,
as $P_1(\bx|\bx\in\Omega)=P_1(\bx)/P_1(\Omega)\approx P_1(\bx)$ for all
$\bx\in\Omega$. This means that the best achievable
compression performance in the sense of (\ref{min}) is roughly the entropy of
$\bX$ under $P_1$, essentially independently of
the choice of $\Omega$, whenever $E_{\mbox{\tiny
MD}}>0$, which makes (\ref{min}) somewhat less interesting than it might seem at first
glance.

It is therefore more interesting to examine objective functions with stronger
sensitivity to the choice of $\Omega$. This would be the case with a large
deviations criterion, like $P_1\{L(\bX)\ge nR|\bX\in\Omega\}$, or the related
criterion of the exponential moment,
$\bE_1[\exp\{\theta L(\bX)\}|\bX\in\Omega]$, where $\theta > 0$ is a
given parameter. These objective functions are not new and they are interesting on their own right 
(see, e.g., \cite[Introduction]{Merhav11} for a discussion on
the motivation). In another version of our problem, we will replace the FA
constraint $P_0(\Omega)\le \exp(-nE_{\mbox{\tiny FA}})$, by a constraint on
the cost of compression in the FA event, namely, a constraint on $P_0\{L(\bX)\ge
nR|\bX\in\Omega\}$ or $\bE_0[\exp\{\theta L(\bX)\}|\bX\in\Omega]$.
In this paper, we focus on these
performance criteria,
as well as on issues of universality, that is,
how to confront uncertainty in $P_0$ and/or $P_1$.
When dealing with these universality issues, we will find
it more convenient to switch the roles between the objective function and one
of the constraints, for example, minimize $P_1(\Omega^c)$ subject to
constraints on $P_0(\Omega)$ and $P_1\{L(\bX)\ge nR|\bX\in\Omega\}$
or on $\bE_1[\exp\{\theta L(\bX)\}|\bX\in\Omega]$.

\section{Preliminaries: A Simple Extension of the Neyman--Pearson Lemma}

The following lemma will turn out to be useful for our purposes
(see also \cite{Merhav13} for a similar lemma).

\begin{lemma}
Let $f$, $g$ and $h$ be any three functions from $\calX^n$ to $\reals$
and let
\begin{equation}
\label{generic}
\Omega_\star=\{\bx:~f(\bx)+a\cdot g(\bx)\le b\cdot h(\bx)\},
\end{equation}
where $a\ge 0$ and $b\ge 0$ are fixed numbers.
Let $\Omega$ be any other subset of $\calX^n$.
If 
\begin{equation}
\sum_{\bx\in\Omega}g(\bx)\le 
\sum_{\bx\in\Omega_\star}g(\bx)
\end{equation}
and
\begin{equation}
\sum_{\bx\in\Omega^c}h(\bx)\le 
\sum_{\bx\in\Omega_\star^c}h(\bx)
\end{equation}
then
\begin{equation}
\sum_{\bx\in\Omega_\star}f(\bx)\le
\sum_{\bx\in\Omega}f(\bx).
\end{equation}
\end{lemma}

The lemma tells us that the decision rule defined by 
$\Omega_\star$ is optimal in the sense that no other 
competing rule $\Omega$ gives strictly smaller values of all three
quantities,
$\sum_{\bx\in\Omega}g(\bx)$, $\sum_{\bx\in\Omega^c}h(\bx)$, and
$\sum_{\bx\in\Omega}f(\bx)$. The paramaters $a$ and $b$ can be thought
of as Lagrange multipliers that control the magnitudes of
$\sum_{\bx\in\Omega_\star}g(\bx)$ and $\sum_{\bx\in\Omega_\star^c}h(\bx)$.  
Note that Lemma 1 (similarly as the classical Neyman--Pearson
lemma) does not require $f$, $g$ and $h$ to be probability
distributions. These can be any functions from $\calX^n$ to $\reals$, in fact, not
necessarily even positive functions. 

{\it Proof.}
Let $\Omega_\star$ be defined as in Theorem  1 and let $\Omega$ be any competing
decision rule.
First, observe that for every $\bx\in\calX^n$
\begin{equation}
[\calI\{\bx\in\Omega_\star\}-\calI\{\bx\in\Omega\}]\cdot\left[b\cdot
h(\bx)-a\cdot g(\bx)-f(\bx)\right]\ge 0.
\end{equation}
This is true because, by definition of $\Omega_\star$,
the two factors of the product at the left--hand side (l.h.s.) are
either both non--positive or both non--negative.
Thus, taking the summation over all $\bx\in\calX^n$, we have:
\begin{equation}
b\cdot\left[\sum_{\bx\in\Omega_\star}h(\bx)-
\sum_{\bx\in\Omega}h(\bx)\right]-a\cdot\left[\sum_{\bx\in\Omega_\star}g(\bx)-
\sum_{\bx\in\Omega}g(\bx)\right]-
\left[\sum_{\bx\in\Omega_\star}f(\bx)-\sum_{\bx\in\Omega}f(\bx)\right]\ge 0
\end{equation}
or, equivalently,
\begin{equation}
\sum_{\bx\in\Omega_\star}f(\bx)-\sum_{\bx\in\Omega}f(\bx)\le
a\cdot\left[\sum_{\bx\in\Omega}g(\bx)-\sum_{\bx\in\Omega_\star}g(\bx)\right]+
b\cdot\left[\sum_{\bx\in\Omega^c}h(\bx)-\sum_{\bx\in\Omega_\star^c}h(\bx)\right].
\end{equation}
Since $a\ge 0$ and $b\ge 0$, then
\begin{equation}
\sum_{\bx\in\Omega}g(\bx)-\sum_{\bx\in\Omega_\star}g(\bx)\le 0
\end{equation}
and
\begin{equation}
\sum_{\bx\in\Omega^c}h(\bx)-\sum_{\bx\in\Omega_\star^c}h(\bx)\le 0
\end{equation}
imply
\begin{equation}
\sum_{\bx\in\Omega_\star}f(\bx)-\sum_{\bx\in\Omega}f(\bx)\le 0,
\end{equation}
which completes the proof of Lemma 1.

\section{Applying Lemma 1 to Joint Detection and Compression}

Lemma 1 is almost applicable for solving one version of 
the problem defined in Section 2. A
simple modification will make it completely applicable. 
First, concerning the constraints, it is
clear that the assignments should be $g(\bx)=P_0(\bx)$ and $h(\bx)=P_1(\bx)$,
for the case of a constraint on $P_0(\Omega)$. 
Regarding the objective function, for a given choice of $\Omega$, the
minimization of $\bE_1\{\exp[\theta L(\bX)|\bX\in\Omega\}$ over all
uniquely decodable length functions, $L(\cdot)$, gives (ignoring integer length
constraints):
\begin{equation}
L^*(\bx)=-\log\left[\frac{[P_1(\bx)]^{1/(1+\theta)}}
{\sum_{\bx'\in\Omega}[P_1(\bx')]^{1/(1+\theta)}}\right],~~~~~\bx\in\Omega
\end{equation}
which yields 
\begin{equation}
\bE_1\{\exp[\theta L^*(\bX)]|\bX\in\Omega\}=
\left(\sum_{\bx\in\Omega}\left[\frac{P_1(\bx)}{P_1(\Omega)}\right]^{1/(1+\theta)}\right)^{1+\theta}.
\end{equation}
Thus, the minimization of $\bE_1\{\exp[\theta L^*(\bX)]|\bX\in\Omega\}$
over $\Omega$ is equivalent to the minimization of
$$\sum_{\bx\in\Omega}\left[\frac{P_1(\bx)}{P_1(\Omega)}\right]^{1/(1+\theta)}.$$
It is tempting now to use Lemma 1 with the additional assignment 
\begin{equation}
\label{illegal}
f(\bx)=\left[\frac{P_1(\bx)}{P_1(\Omega)}\right]^{1/(1+\theta)},
\end{equation}
but this is not quite a legitimate choice for using Lemma 1, 
since this function depends on $\Omega$.
Nonetheless, as observed in Section 2, in the regime where $P_1(\Omega^c)\ge
1-\exp(-nE_{\mbox{\tiny MD}})\to 1$, the factor $P_1(\Omega)$ has an
asymptotically negligible effect, and we can uniformly approximate by choosing
\begin{equation}
f(\bx)=[P_1(\bx)]^{1/(1+\theta)}.
\end{equation}
Also, in order for the coefficients $a$ and $b$ to influence the asymptotic
exponents of the objective function and the constraints, we let 
them be exponential functions of $n$, i.e., $a=e^{n\alpha}$ and
$b=e^{n\beta}$, where $\alpha$ and $\beta$ are fixed real numbers, independent
of $n$, which are dictated by $E_{\mbox{\tiny FA}}$ and $E_{\mbox{\tiny MD}}$.
The asymptotically optimal decision rule now reads
\begin{equation}
\label{omegastar}
\Omega_\star=\{\bx:~[P_1(\bx)]^{1/(1+\theta)}+e^{n\alpha} P_0(\bx)\le
e^{n\beta}
P_1(\bx)\}.
\end{equation}

\section{Discussion and Analysis of the Decision Rule}

Let us now examine the decision rule $\Omega_\star$, defined in eq.\
(\ref{omegastar}). Since
\begin{eqnarray}
\max\left\{[P_1(\bx)]^{1/(1+\theta)}, e^{n\alpha}P_0(\bx)\right\}
&\le&[P_1(\bx)]^{1/(1+\theta)}+e^{n\alpha}P_0(\bx)\\
&\le&
2\cdot\max\left\{[P_1(\bx)]^{1/(1+\theta)}, e^{n\alpha} P_0(\bx)\right\},
\end{eqnarray}
the performance of $\Omega_\star$ is asymptotically equivalent (in terms of
asymptotic exponents of the objective function, the FA probability and the MD
probability)
to that of
\begin{eqnarray}
\label{omegahat}
\hat{\Omega}&\dfn&\left\{\bx:~\max\{[P_1(\bx)]^{1/(1+\theta)},e^{n\alpha}
P_0(\bx)\}\le e^{n\beta}
P_1(\bx)\right\}\nonumber\\
&=&\{\bx:~[P_1(\bx)]^{1/(1+\theta)}\le
e^{n\beta}P_1(\bx),~e^{n\alpha} P_0(\bx)\le e^{n\beta}P_1(\bx)\}\nonumber\\
&=&\left\{\bx:~-\ln P_1(\bx)\le n\beta\left(1+\frac{1}{\theta}\right),~
\ln\left[\frac{P_1(\bx)}{P_0(\bx)}\right]\ge n(\alpha-\beta)\right\}
\end{eqnarray}
The form of $\hat{\Omega}$ is more convenient than that of 
$\Omega_\star$, both for understanding the
behavior, and for implementation (since it allows passage to the logarithmic
domain as is shown in the last line of eq.\ (\ref{omegahat})).
The test $\hat{\Omega}$ can be thought of as a combination of two tests: (i) the test
$-\ln P_1(\bx)\le n\beta(1+1/\theta)$, which guarantees that the code--length
associated with $\bx$ is small enough, and (ii) the test
$\ln[P_1(\bx)/P_0(\bx)]\ge n(\alpha-\beta)$, which is the
ordinary likelihood ratio test that distinguishes between $P_0$ and $P_1$.
The test $\hat{\Omega}$ also lends itself more conveniently to standard
asymptotic exponent
analysis using the method of types \cite{CK81}. The results are as follows.

Consider the MD probability first.
\begin{equation}
P_1(\Omega_\star^c)\exe P_1(\hat{\Omega}^c)\exe 
\exp\{-ne_{\mbox{\tiny MD}}\}
\end{equation}
where
\begin{eqnarray}
e_{\mbox{\tiny MD}}&=&\min_Q\{\calD(Q\|P_1):~\bE_Q\ln P_1(X)\le
-\beta(1+1/\theta)~\mbox{or}~\bE_Q\ln[P_1(X)/P_0(X)]\le \alpha-\beta\}\\
&=&\min\{e_1(\beta),e_2(\alpha-\beta)\}
\end{eqnarray}
with
\begin{equation}
e_1(\beta)=\min_Q\{\calD(Q\|P_1):~\bE_Q\ln P_1(X)\le
-\beta(1+1/\theta)\}
\end{equation}
and 
\begin{equation}
e_2(\alpha-\beta)=\min_Q\{\calD(Q\|P_1):~\bE_Q\ln[P_1(X)/P_0(X)]\le \alpha-\beta\}.
\end{equation}
Here $\bE_Q\{\cdot\}$ denotes the expectation operator w.r.t.\ a generic
probability distribution $Q$ on $\calX$ and $\calD(Q\|P)$
is the Kullback--Leibler divergence between $Q$ and $P$. 
Both $e_1(\beta)$ and $e_2(\alpha-\beta)$ must be no smaller than $E_{\mbox{\tiny MD}}$.
Both minimization problems can easily be solved using Lagrange multipliers.
The minimizing $Q$ for $e_1$ is of the form
\begin{equation}
Q_1(x)=\frac{[P_1(x)]^\lambda}{\sum_{x'\in\calX}[P_1(x')]^\lambda},~~~~
\lambda\le 1
\end{equation}
where $\lambda$ is chosen to satisfy the constraint $~\bE_Q\ln P_1(X)\le
-\beta(1+1/\theta)$. Clearly, $e_1(\beta)$ is a monotonically increasing
function, and due to its convexity, strictly so in the range
where it is non--zero and finite, which is 
$\theta H_1/(1+\theta)< \beta\le -\theta[\ln\min_xP_1(x)]/(1+\theta)$, $H_1$ being the
entropy of $P_1$. Thus, we must choose $\beta\ge e_1^{-1}(E_{\mbox{\tiny
MD}})$. Similarly, the minimzing $Q$ for $e_2$ is of the form
\begin{equation}
Q_2(x)=\frac{[P_0(x)]^\nu [P_1(x)]^{1-\nu}}{Z(\nu)}
\end{equation}
where $\nu\ge 0$ is chosen to satisfy the constraint $\bE_Q\ln[P_1(X)/P_0(X)]\le
\alpha-\beta$ and $Z(\nu)$ is a normalization constant. 
The convex function $e_2$ is strictly decreasing in $\alpha-\beta$
in the range where it is positive and finite, 
$\min_x\ln[P_1(x)/P_0(x)]\le \alpha-\beta< \calD(P_1\|P_0)$.
Thus, we must choose $\alpha-\beta \le e_2^{-1}(E_{\mbox{\tiny
MD}})$. Clearly, once  we have selected some $\beta\ge e_1^{-1}(E_{\mbox{\tiny
MD}})$, the best choice of $\alpha$ (that would maximally shrink
$\Omega_\star$, or $\hat{\Omega}$) would be the maximum allowed value, 
$\alpha=\beta + e_2^{-1}(E_{\mbox{\tiny
MD}})$. The choice of $\beta$ will then be dictated by the FA constraint.
This simple observation reduces the original space of trade-offs with two degrees of
freedom ($\alpha$ and $\beta$) to one degree of freedom ($\beta$ only).

Concerning the FA probability, we have
\begin{equation}
P_0(\Omega_\star)\exe P_0(\hat{\Omega})\exe 
\exp\{-ne_{\mbox{\tiny FA}}\}
\end{equation}
where
\begin{eqnarray}
e_{\mbox{\tiny FA}}&=&\min_Q\{\calD(Q\|P_0):-\bE_Q\ln P_1(X)\le
\beta(1+1/\theta),\bE_Q\ln[P_0(X)/P_1(X)]\le \beta-\alpha\}\nonumber\\
&=&\min_Q\{\calD(Q\|P_0):-\bE_Q\ln P_1(X)\le
\beta(1+1/\theta),\bE_Q\ln[P_0(X)/P_1(X)]\le -e_2^{-1}(E_{\mbox{\tiny MD}})\}.\nonumber
\end{eqnarray}
Similarly as before,
the minimizing $Q$,
denoted $Q^*$, is of the form
\begin{equation}
Q^*(x)=\frac{[P_0(x)]^{1-\eta}[P_1(x)]^{\eta+\xi}}{Z(\eta,\xi)},
\end{equation}
where $Z(\eta,\xi)$ is a normalization constant, and where
$\eta\ge 0$ and $\xi\ge 0$ are chosen to satisfy the constraints,
$-\bE_Q\ln P_1(X)\le
\beta(1+1/\theta)$ and $\bE_Q\ln[P_0(X)/P_1(X)]\le -e_2^{-1}(E_{\mbox{\tiny
MD}})$. Here, $e_{\mbox{\tiny FA}}$ is a decreasing function of $\beta$, and
so, the constraint $e_{\mbox{\tiny FA}}(\beta)\ge E_{\mbox{\tiny FA}}$
dictates the choice $\beta\le e_{\mbox{\tiny FA}}^{-1}(E_{\mbox{\tiny FA}})$,
which is feasible (in view of the earlier MD exponent analysis)
provided that $e_{\mbox{\tiny FA}}^{-1}(E_{\mbox{\tiny FA}})\ge
e_1^{-1}(E_{\mbox{\tiny MD}})$. Under this condition, it is possible to assign
\begin{equation}
\label{alpha}
\alpha=e_{\mbox{\tiny FA}}^{-1}(E_{\mbox{\tiny FA}})+e_2^{-1}(E_{\mbox{\tiny
MD}}) 
\end{equation}
and 
\begin{equation}
\label{beta}
\beta=e_{\mbox{\tiny FA}}^{-1}(E_{\mbox{\tiny FA}}).
\end{equation}
Finally, using the method of types once again,
the exponent associated with the objective function is given by
\begin{equation}
\bE_1[\exp\{\theta L^*(\bX)\}|\bX\in\Omega_\star]=
\left(\sum_{\bx\in\hat{\Omega}}[P_1(\bx)]^{1/(1+\theta)}\right)^{1+\theta}\\
\exe \exp\{ne_c\},
\end{equation}
where
\begin{equation}
e_c=\max_Q\{\theta H(Q)-\calD(Q\|P_1):~\bE_Q\ln P_1(X)\ge
-\beta(1+1/\theta),~\bE_Q\ln[P_1(X)/P_0(X)]\ge \alpha-\beta\},
\end{equation}
with $\alpha$ and $\beta$ as in eqs.\ (\ref{alpha}) and (\ref{beta}), and
with $H(Q)$ being the entropy associated with a distribution $Q$ on $\calX$. 
Once again, this is a convex programming problem that 
can be solved using Lagrange multipliers. This completes the analysis of
asymptotic exponents associated with $\Omega_\star$.

As $\alpha$, $\beta$ and $\theta$ vary, it is expected that these exponents
may exhibit certain phase transitions, because of possible abrupt passages between
regions where one of the constraints is active to regions where the other one
becomes active (or both). The following is a simple example
that demonstrates this point.

\noindent
{\it Example.} Let $\calX=\{0,1\}$, define
$P_0$ to be the binary symmetric source (BSS) and let $P_1$ be defined by
$P_1(1)=1-P_1(0)=3/4$. In this case, 
it is straightforward to verify that $\hat{\Omega}$ is the set of all source
vectors $\{\bx\}$ for
which the relative frequency of 1's is at least as large as
\begin{equation}
\label{qab}
q_{\alpha,\beta}=\frac{\max\{\ln 4-\beta(1+1/\theta),\ln 2+\alpha-\beta\}}{\ln
3}.
\end{equation}
As long as $q_{\alpha,\beta}\in (1/2,3/4)$,
the error exponents are simply
\begin{equation}
e_{\mbox{\tiny FA}}=D\left(q_{\alpha,\beta}\|\frac{1}{2}\right),~~~
e_{\mbox{\tiny MD}}=D\left(q_{\alpha,\beta}\|\frac{3}{4}\right),
\end{equation}
where for $s,t\in[0,1]$, $D(s\|t)$ denotes the binary
divergence, i.e., $D(s\|t)=s\ln(s/t)+(1-s)\ln[(1-s)/(1-t)]$. 
It is assumed, of course, that $E_{\mbox{\tiny FA}}$ and $E_{\mbox{\tiny
MD}}$ are small enough such that there exist 
$\alpha$ and $\beta$ with
$D(q_{\alpha,\beta}\|\frac{1}{2})\ge E_{\mbox{\tiny FA}}$
and $D(q_{\alpha,\beta}\|\frac{3}{4})\ge E_{\mbox{\tiny MD}}$. 
The derivatives of the exponents
$e_{\mbox{\tiny FA}}$ and $e_{\mbox{\tiny MD}}$,
as functions of $\alpha$, $\beta$
and $\theta$, are discontinuous at the points where
\begin{equation}
\ln 4-\beta\left(1+\frac{1}{\theta}\right)=\ln 2+\alpha-\beta, 
\end{equation}
because at these points, the
achiever of the maximum on the r.h.s.\ of (\ref{qab}) switches between
the two arguments of
the $\max$ operator.
These are therefore points of phase transitions.

\section{Other Variants of the Problem}

As mentioned in Section 2, it makes sense to 
replace the FA constraint by a
constraint that quantifies the true cost of the FA error, namely, superfluous data compression.
This suggests to replace $g(\bx)=P_0(\bx)$ by 
$g(\bx)=P_0(\bx)e^{\theta L^*(\bx)}$,
where $L^*(\bx)$ is still defined as above because when $\bx\in\Omega$, we
believe that the underlying source is $P_1$. 
This amounts to 
\begin{equation}
g(\bx)=P_0(\bx)[P_1(\bx)]^{-\theta/(1+\theta)}
\left(\sum_{\bx'\in\Omega}[P_1(\bx')]^{1/(1+\theta)}\right)^\theta.
\end{equation}
The problem is that now, similarly as in (\ref{illegal}), Lemma 1
is not directly applicable since $g$ depends on $\Omega$ and in a
non--trivial manner. 

There is, however, a way to circumvent this difficulty, that both improves performance and
allows to use Lemma 1. Let us replace $L^*(\bx)$ by the length function of a
{\it universal} encoder, which will be nearly optimal no matter 
whether $P_0$ or $P_1$ (or any other memoryless source) is the
true underlying source. The best we can do is use a universal code whose
length function, $L_U(\bx)$, is essentially as small as $n\hat{H}_{\bx}(X)$ (up to
a sub-linear additional term),
where $\hat{H}_{\bx}(X)$ stands for the empirical entropy of $\bx$, namely,
the entropy associated with the empirical distribution of $\bx$.\footnote{For example,
consider a
two--part code that first describes the index of the type class and then the
location of $\bx$ within the type class.} Such a code is known to be
asymptotically optimal, not only in the sense of the expected code--length,
but also for
a very wide class of additional criteria (see \cite{WMF94}), 
including $\bE_1\exp\{\theta L(\bX)|\bX\in\Omega\}$ and
$P_1\{L(\bX)\ge n|\bX\in\Omega\}$.\footnote{The fact that $L_U(\bx)$ asymptotically achieves 
the minimum of $\bE_1\exp\{\theta L(\bx)|\bX\in\Omega\}$, which is
approximated by
$[\sum_{\bx\in\Omega}[P_1(\bx)]^{1/(1+\theta)}]^{1+\theta}$,
can easily be verified using the method of types. Concerning the criterion 
$P_1\{L(\bX)\ge nR|\bX\in\Omega\}$, it achieves an error exponent of
$\min\{\calD(Q\|P_1):~H(Q)\ge R,~\calT(Q)\subseteq\Omega\}$, 
which is the best possible, as can easily be
shown by a straightforward modification of the converse part of
\cite[Theorem 1]{Marton74}.}
We can now apply Lemma 1 with the choice
\begin{equation}
\label{g}
g(\bx)=P_0(\bx)\exp\{n\theta\hat{H}_{\bx}(X)\}.
\end{equation}
By the same token, the choice of $f$ can also be changed to
\begin{equation}
\label{f}
f(\bx)=P_1(\bx)\exp\{n\theta\hat{H}_{\bx}(X)\}.
\end{equation}
More generally, one can use, of course, two different values of $\theta$, say, $\theta_0$
and $\theta_1$ in eqs.\ (\ref{g}) and (\ref{f}), 
respectively, and finally, re--define $\Omega_\star$ accordingly to read
\begin{equation}
\label{newomegastar}
\Omega_\star=\left\{\bx:~P_1(\bx)\exp\{n\theta_1\hat{H}_{\bx}(X)\}+
e^{n\alpha}P_0(\bx)\exp\{n\theta_0\hat{H}_{\bx}(X)\}\le
e^{n\beta}P_1(\bx)\right\}.
\end{equation}
Similarly, 
we can now address directly the excess code--length criterion
by choosing
\begin{equation}
g(\bx)=P_0(\bx)\cdot\calI\{\bx:~\hat{H}_{\bx}(X)\ge R\},
\end{equation}
\begin{equation}
f(\bx)=P_1(\bx)\cdot\calI\{\bx:~\hat{H}_{\bx}(X)\ge R\},
\end{equation}
and again, define $\Omega_\star$ accordingly.
It should be emphasized that this passage from $L^*(\bx)$ to
$n\hat{H}_{\bx}(X)$ is not accompanied by loss in performance in terms of
asymptotic exponents.

In the case of lossy source coding, $\hat{H}_{\bx}(X)$, throughout this discussion, should be
replaced by the empirical rate--distortion function, namely, the
rate-distortion function associated with the empirical distribution induced by
$\bx$, or the empirical distortion--rate function, depending on the assumed
regime, fixed--distortion and minimum rate or vice versa (see \cite{AM98}). 
In all these variants, the asymptotic exponential performance can easily be
assessed using the method types, similarly as before. 

\section{Universal Decision Rules}

In the previous section, we discussed the use of universal lossless source
coding, which facilitates the use of Lemma 1, and at the same time, makes sense even if $P_0$
and $P_1$ are known, because when $\bx\in\Omega$, there is never full
certainty that it has really emerged from $P_1$. But what happens if $P_0$ and $P_1$
are not both known (except for being memoryless)? The latest proposed version of $\Omega_\star$
(eq.\ (\ref{newomegastar})) still depends on $P_0$ and $P_1$, and hence not
implementable in this case.
We next turn to handle universality issues associated with the choice of the
decision rule. 
The methodology here is similar to that of a few
earlier papers on universal hypothesis testing (see, e.g., 
\cite{Merhav00}, \cite{MGZ89}, \cite{ZZM92}, \cite{Ziv88a},
\cite{Ziv88b}). Lemma 1 is no longer used explicitly. 

As a starting point, it will be more convenient to consider the problem 
\begin{eqnarray}
\label{minuniv}
& &\mbox{minimize}~~~P_1(\Omega^c)\nonumber\\
& &\mbox{subject to}~~P_0(\Omega)\le e^{-nE_{\mbox{\tiny FA}}}
\nonumber\\
& &~~~~~~~~~~~~~~~\sum_{\bx\in\Omega}
P_1(\bx)\exp\{n\theta_1\hat{H}_{\bx}(X)\}\le e^{\lambda_1 n},
\end{eqnarray}
which is equivalent to one of the versions of the earlier problem, except that
the objective function and one of the constraints have interchanged their roles.

We begin with the case where $P_0$ is known but
$P_1$ is not.
Since $P_1$ is unknown, the second constraint must be imposed for {\it every} memoryless
source $P_1$, that is, 
\begin{equation}
\max_{P_1}\sum_{\bx\in\Omega} P_1(\bx)\exp\{n\theta_1\hat{H}_{\bx}(X)\}\le
e^{\lambda_1 n}.
\end{equation}
First, observe that
without loss of
asymptotic optimality, every type class of source vectors, $\calT_{\bx}$, can
be assumed to
belong in its entirety to either $\Omega$ or $\Omega^c$.\footnote{If this is
not the case, then at least half of the members of the type class belong to either $\Omega$
or $\Omega^c$. By transferring the smaller part of each type class to the other
decision region, to join the majority therein, 
one at most doubles the probability of that region, while
reducing the probability of the other region. This has
no negative impact on the asymptotic exponents.}
Accordingly, let $\calT_{\bx}\subseteq \Omega$. Then,
\begin{eqnarray}
e^{\lambda_1 n}&\ge&\max_{P_1}\sum_{\bx\in\Omega}
P_1(\bx)\exp\{n\theta_1\hat{H}_{\bx}(X)\}\\
&\ge&\max_{P_1}\sum_{\bx'\in\calT_{\bx}}
P_1(\bx')\exp\{n\theta_1\hat{H}_{\bx'}(X)\}\\
&=&\max_{P_1}|\calT_{\bx}|\cdot P_1(\bx) \exp\{n\theta_1\hat{H}_{\bx}(X)\}\\
&=&\exp\{n\theta_1\hat{H}_{\bx}(X)-O(\log n)\}.
\end{eqnarray}
The conclusion is then that $\calT_{\bx}\subseteq \Omega$ implies
$\calT_{\bx}\subseteq \{\bx:~\hat{H}_{\bx}(X)\le \lambda_1/\theta_1+O(\log n/n\}$, which
means
\begin{equation}
\Omega\subseteq \{\bx:~\hat{H}_{\bx}(X)\le \lambda_1/\theta_1+O(\log n/n)\}.
\end{equation}
From the first constraint of (\ref{minuniv}), we similarly have:
\begin{equation}
\label{div}
\Omega\subseteq \{\bx:~\calD(\hat{P}_{\bx}\|P_0)\ge E_{\mbox{\tiny FA}}-O(\log
n/n)\},
\end{equation}
where $\hat{P}_{\bx}$ is the empirical distribution associated with $\bx$.
Combining the last two equations, we get:
\begin{equation}
\label{include}
\Omega\subseteq \Omega_u\dfn\{\bx:~\hat{H}_{\bx}(X)\le
\lambda_1/\theta_1+O(\log n/n),~\calD(\hat{P}_{\bx}\|P_0)\ge E_{\mbox{\tiny
FA}}-O(\log n/n)\}.
\end{equation}
We now propose $\Omega_u$ as our universal decision rule. First, observe that
it asymptotically satisfies the constraints, as
\begin{equation}
P_0(\Omega_u)\le P_0\{\bx:~\calD(\hat{P}_{\bx}\|P_0)\ge E_{\mbox{\tiny
FA}}-O(\log n/n)\}\exe \exp\{-n[E_{\mbox{\tiny FA}}-O(\log n/n)\},
\end{equation}
and
\begin{eqnarray}
\sum_{\bx\in\Omega_u}P_1(\bx)\exp\{n\theta_1\hat{H}_{\bx}(X)\}&\le&
\max_{P_1}\sum_{\bx\in\Omega_u}P_1(\bx)\exp\{n\theta_1\hat{H}_{\bx}(X)\}\nonumber\\
&\le&\sum_{\bx\in\Omega_u}\max_{P_1}P_1(\bx)\exp\{n\theta_1\hat{H}_{\bx}(X)\}\nonumber\\
&\le&\sum_{\bx\in\Omega_u}\exp\{-n\hat{H}_{\bx}(X)\}\cdot
\exp\{n\theta_1\hat{H}_{\bx}(X)\}\nonumber\\
&=&\sum_{\calT_{\bx}\subset \Omega_u}|\calT_{\bx}|\cdot\exp\{-n\hat{H}_{\bx}(X)\}\cdot
\exp\{n\theta_1\hat{H}_{\bx}(X)\}\nonumber\\
&\exe&\max_{\calT_{\bx}\subset \Omega_u}
\exp\{n\theta_1\hat{H}_{\bx}(X)\}\nonumber\\
&\exe& e^{\lambda_1 n}.
\end{eqnarray}
On the other hand, since $\Omega_u$ is a super-set of any competing decision
rule $\Omega$ that satisfies the constraints (see eq.\ (\ref{include})), then it follows that
$\Omega_u^c\subseteq \Omega^c$, and so, $P_1(\Omega_u^c)\le P_1(\Omega^c)$,
for every $P_1$. This means that $\Omega_u$ minimizes the MD probability
{\it uniformly} for every memoryless source $P_1$ and hence
establishes the optimality of $\Omega_u$.

The idea here is that $\Omega_u$ is essentially the largest subset of $\calX^n$ that still
satisfies the constraints, and hence its complement is the smallest possible.
Once again, we see that membership in $\Omega_u$ consists of two requirements:
the requirement on the empirical entropy, which limits the code length, and a
requirement on the divergence, which means that $\bx$ is far enough from being
typical to $P_0$, in order to reject unwanted data that stems from $P_0$.

Universal counterparts of other variants of the problem, discussed in the
previous section, can be derived in a
similar manner. For example,
if the constraint $P_0(\Omega)\le e^{-nE_{\mbox{\tiny FA}}}$ is replaced by
compression cost constraint
\begin{equation}
\sum_{\bx\in\Omega}P_0(\bx)\exp\{n\theta_0\hat{H}_{\bx}(X)\}\le e^{\lambda_0 n}
\end{equation}
then the set $\{\bx:~\calD(\hat{P}_{\bx}\|P_0)\ge E_{\mbox{\tiny FA}}-O(\log
n/n)\}$, in eq.\ (\ref{div}), should be
replaced by $\{\bx:~\theta_0\hat{H}_{\bx}(X)-\calD(\hat{P}_{\bx}\|P_0)\le
\lambda_0\}$ and $\Omega_u$ should, of course, be modified accordingly.
If, in addition, $P_0$ is unknown as well, and this constraint is imposed for every
memoryless source $P_0$ on $\calX$, then
this becomes
$\{\bx:~\theta_0\hat{H}_{\bx}(X)\le
\lambda_0\}$. Thus, overall $\Omega_u$ would be redefined as
\begin{equation}
\Omega_u=\{\bx:~\hat{H}_{\bx}(X)\le\min\{\lambda_0/\theta_0,\lambda_1/\theta_1\}\}.
\end{equation}

Suppose next that both $P_0$ and $P_1$ are unknown but there are training sequences
available from each one of these sources. In other words, in addition to the vector
$\bx\in\calX^n$ as before, we also have a training sequence,
$\bx_0\in\calX^m$ from $P_0$,
and a training sequence, $\bx_1\in\calX^m$ from $P_1$.
A natural approach would be the plug--in approach: First estimate each source
from its corresponding training data and then use each estimate in place of the
corresponding unknown, true source. This is a sub-optimal approach because it is based on
separation and it does not use $\bx$ for estimating either source.
The best approach is to combine the training
and the decision into a single step, which means that our decision rule
classifies triples $\{(\bx,\bx_0,\bx_1)\}$ rather than single vectors
$\{\bx\}$ as before. The compression cost constraints will now
read 
\begin{equation}
\sum_{\bx,\bx_0,\bx_1\in\Omega}
P_0(\bx_0)P_1(\bx_1)P_0(\bx)\exp\{n\theta_0\hat{H}_{\bx}(X)\}\le
\exp(\lambda_0 n)
\end{equation}
and
\begin{equation}
\sum_{\bx,\bx_0,\bx_1\in\Omega}
P_0(\bx_0)P_1(\bx_1)P_1(\bx)\exp\{n\theta_1\hat{H}_{\bx}(X)\}\le
\exp(\lambda_1 n),
\end{equation}
both imposed for every two memoryless sources $P_0$ and $P_1$.
Here, we assume, again without loss of asymptotic optimality, that
$\Omega$ is a union of Cartesian products of type classes
$\calT_{\bx}\times\calT_{\bx_0}\times\calT_{\bx_1}$.
As for the first constraint, we have
\begin{eqnarray}
e^{\lambda_0
n}&\ge&\max_{P_0,P_1}\sum_{\bx,\bx_0,\bx_1\in\Omega}
P_0(\bx_0)P_1(\bx_1)P_0(\bx)\exp\{n\theta_0\hat{H}_{\bx}(X)\}\\
&\ge& \exp\{n\theta_0\hat{H}_{\bx}(X)-n\calD(\hat{P}_{\bx}\|\hat{P}_{\bx\bx_0})-m
\calD(\hat{P}_{\bx_0}\|\hat{P}_{\bx\bx_0})-O(\log n)\},
\end{eqnarray}
where $\hat{P}_{\bx\bx_i}$ denotes the empirical distribution associated with
the concatenation of $\bx$ and $\bx_i$, $i=0,1$.
Similarly, for the other constraint
\begin{eqnarray}
e^{\lambda_1
n}&\ge&\max_{P_0,P_1}\sum_{\bx,\bx_0,\bx_1\in\Omega}
P_0(\bx_0)P_1(\bx_1)P_1(\bx)\exp\{n\theta_1\hat{H}_{\bx}(X)\}\\
&\ge&
\exp\{n\theta_1\hat{H}_{\bx}(X)-n\calD(\hat{P}_{\bx}\|\hat{P}_{\bx\bx_1})-m
\calD(\hat{P}_{\bx_1}\|\hat{P}_{\bx\bx_1})-O(\log n)\}
\end{eqnarray}
and then $\Omega_u$ is defined as
\begin{eqnarray}
\Omega_u&=&\{\bx:~\theta_0\hat{H}_{\bx}(X)-\calD(\hat{P}_{\bx}\|\hat{P}_{\bx\bx_0})-\frac{m}{n}
\calD(\hat{P}_{\bx_0}\|\hat{P}_{\bx\bx_0})-O(\log n/n)\le\lambda_0,\nonumber\\
& &\theta_1\hat{H}_{\bx}(X)-\calD(\hat{P}_{\bx}\|\hat{P}_{\bx\bx_1})-\frac{m}{n}
\calD(\hat{P}_{\bx_1}\|\hat{P}_{\bx\bx_1})-O(\log n/n)\le\lambda_1\}.
\end{eqnarray}
The terms $\calD(\hat{P}_{\bx}\|\hat{P}_{\bx\bx_i})+\frac{m}{n}
\calD(\hat{P}_{\bx_1}\|\hat{P}_{\bx\bx_i})$ measure the `distance' between
$\hat{P}_{\bx}$ and $\hat{P}_{\bx_i}$, $i=0,1$. If they are close, these terms
are small and we compare the code length to a threshold. If they are far
apart, we can afford to be more tolerant concerning the length since this is a rare event
anyway. The empirical distributions $\hat{P}_{\bx\bx_0}$ and
$\hat{P}_{\bx\bx_1}$ stand for the fact that, in some sense, $\bx$ participates in the
estimation of the two sources, unlike the `plug-in' approach describe above.

\section{Conclusion}

We have addressed the problem of joint detection and lossless data compression
in several variants, including the universal regime, where at least one of the
sources is unknown, with and without training sequences from each source.
The method of our derivations can also be carried out in several more general
situations.

First, it is not difficult to
extend our results from memoryless sources to Markov sources, or even more generally,
to unifilar finite--state sources. This is possible because the method of types
extends to these classes of sources as well. Moreover, in the universal
setting, it is more interesting to consider the case where the Markov order is
unknown (or in the case of unifilar finite--state sources, the
state--transition diagram and the number of states are unknown). In this case,
it is expected that the length function of the 
Lempel--Ziv algorithm can be invoked instead of the empirical entropy, similarly as was
done in earlier work (see, e.g., \cite{Merhav00}, \cite{MGZ89}).

Secondly, as mentioned already in the Abstract and the
Introduction, one may consider tasks other than lossless data compression.
One of them is lossy data compression, and we have already mentioned, at the end
of Section 6, how to modify our decision rule to account for this case.
Channel decoding is another important task that 
has already been addressed in \cite{Merhav13}. Additional
tasks may be quantization, estimation, encryption, and so on.
The general guideline is always to try to present (or approximate) the objective function
(pertaining to the optimal strategy of the task within $\Omega$) as
(a monotonic function of) the summation or integral of some function $f(\bx)$ over
$\Omega$, and then use this $f$ in the decision rule $\Omega_\star$
of eq.\ (\ref{generic}). The function $f$ should be independent of $\Omega$.



\end{document}